\documentstyle[12pt]{article}
\begin{document}

\renewcommand{\baselinestretch}{1.5}
{\noindent \small \it Eur. Phys. J. B --- rapid communication}
\vskip 1.0cm

\begin{center} {\Large \bf A New Universality}\\ $\ $\\
{\Large \bf for }\\ $\ $ \\
{\Large \bf Random Sequential Deposition of Needles}

\vskip 0.6cm {\bf N.Vandewalle$^{1,2}$, S.Galam$^2$ and M.Kramer$^{1,3}$}

\vskip 0.6cm
$^1$ GRASP, Institut de Physique B5, Universit\'e de Li\`ege, \\ B-4000
Li\`ege, Belgium.
\vskip 0.6cm

$^2$ Laboratoire des Milieux D\'esordonn\'es et H\'et\'erog\`enes, \\ Tour
13, Case 86, 4 place Jussieu, 75252 Paris Cedex 05, France.
\vskip 0.6cm

$^3$ 2, rue Kockelberg, L-9252 Diekirch, Luxembourg.

\end{center}

\vskip 0.6cm


{\noindent \large Abstract}
\vskip 0.6cm

Percolation and jamming phenomena are investigated for random sequential
deposition of rectangular needles on $d=2$ square lattices. Associated
thresholds $p_c^{perc}$ and $p_c^{jam}$ are determined for various needle
sizes. Their ratios $p_c^{perc} / p_c^{jam}$ are found to be a constant
$0.62 \pm 0.01$ for all sizes. In addition the ratio of jamming thresholds
for respectively square blocks and needles is also found to be a constant
$0.79 \pm 0.01$. These constants exhibit some universal connexion in the
geometry of jamming and percolation for both anisotropic shapes (needles
versus square lattices) and isotropic shapes (square blocks on square
lattices). A universal empirical law is proposed for all three thresholds
as a function of $a$.

\newpage

Percolation phenomena are generic in the study of disordered media like
porous materials, alloys, ecosystems, etc... \cite{bookstauffer}. A
percolation transition is based on calculating the probability of
occurrence of an infinite connectivity between elements of the random
medium as a function of the fraction $p$ of constitutive elements. The
usually concerned object is the percolating cluster connecting distant
borders of the medium. At the critical point $p_c^{perc}$, this cluster is
very tortuous and is a fractal object. Even though some advances have been
made in the understanding of percolation, numerous questions remains
unanswered today \cite{percoquest}. Among others, the exact values of many
thresholds are still not known though some powerful conjectures have been
proposed for a rather large class of lattices \cite{gm}.

Beside connectivity properties, another fundamental question related to
disordered structures is the construction of randomly packed structures.
Among other models, the random deposition also called Random Sequential
Adsorption (RSA), leads to a critical phenomenon. Above a critical fraction
$p_c^{jam}$ of occupied sites, the medium becomes fully saturated. No more
object can be inserted in it \cite{rsa}. At this stage the packing density
of (identical) objects filling nearly the whole medium is at maximum.

In this paper, we investigate both percolation and jamming phenomena of
non-overlapping anisotropic objects (needles) on strictly $d=2$ square
lattices, needles having commensurate scales with respect to the lattice
spacing. The study of packed anisotropic objects is relevant for
geometrical--physical properties of granular media like the electrical
properties of metallic needles \cite{pirard} or the compaction properties
of granules \cite{tetris}. We will see that both phenomena (percolation and
jamming) become highly related to each other independently of the aspect
ratio $a$ of the needles.

Earlier studies of needles within the percolation framework have been
investigated in the overlapping case only and have been introduced for e.g.
mimicking the formation of microfractures in a brittle material
\cite{robinson}. The percolation of non-overlapping needles has never been
studied to our knowledge. The RSA of needles has only been considered for
specific aspect ratio values, e.g. $a=2$ on $d=2$ lattices \cite{nord}, and
for off-lattice cases \cite{vigil}. It seems that the RSA of needles on
discrete lattices has not been considered up to now. The RSA of (isotropic)
blocks on a square lattice has been numerically investigated by Nakamura
\cite{nakamura} and thereafter confirmed in \cite{privman}. We will also
compare our work to Nakamura's result in order to emphasize the effect of
anisotropy. Also, an unexpected needle-block relationship will be
emphasized in the following!

On the square lattice, touching needles are needles which have at least one
cell side in common. Figure 1 illustrates both percolation and jamming
clusters for the particular aspect ratio $a=4$. The figure presents a $16
\times 16$ square lattice at $p_c^{perc}$ and $p_c^{jam}$ for the case of
$4 \times 1$ needles, and a jammed phase for $4 \times 4$ blocks. In the
case of needles, one observes that both transitions take place at quite
different values of the fraction $p$ of occupied sites. As we will see
below, the percolation threshold is ``low" implying that the unoccupied
sites (holes) can form open and large pores. At the percolation threshold,
the needle structure is highly heterogeneous. When the fraction of needles
increases, the large pores are filled up to the jamming situation. Near the
jamming transition, some short range order appears. Indeed, a close packing
of needles oriented in the same direction is observed. Close (oriented)
packed needles seem to form blocks of size $a \times a$! In order to
emphasize this ordering, a typical jamming phase is presented in Figure 2
for $L=400$ and for $a=20$. Horizontal and vertical orientations of needles
are represented in different grey levels. The different colors put into
evidence the short range ordering. More precisely, some domains of
horizontal or vertical needles are seen.

In the third case illustrated in Figure 1, $4 \times 4$ blocks are seen in
the jamming phase. Large spanning clusters of connected blocks are not
formed. One should note that percolation does not occur below the jamming
threshold in the case of blocks. However, percolation always takes place
before jamming in the case of needles. This makes a big difference between
the deposition of anisotropic and isotropic non-overlapping objects on
square lattices.

In order to determine the thresholds for both phenomena, numerical
simulations have been performed on $d=2$ square lattices containing up to
$2000 \times 2000$ sites. Starting from an empty lattice, needles are added
sequentially such that $p$ increases linearly. The percolation and jamming
phenomena are checked until they are found. The first percolation
threshold, i.e. when two distant borders are connected, is considered here,
in contrast to the second percolation threshold which is obtained when the
four borders are connected \cite{ak}. The probability $P_{perc}$ to find a
percolating cluster and the probability $P_{jam}$ to find a jamming phase
are fitted by the error function \begin{equation} P = {1 \over \sqrt{2\pi}
\Delta}\int_{-\infty}^p \exp{\left[ - {1 \over 2} {\left( {p'-p_c \over
\Delta} \right)}^2 \right]} dp' \end{equation} where $p_c$ is the critical
point and $\Delta$ is the width of the transition. Our assumption that the
distribution of critical points is a Gaussian is sufficient from a
practical point of view, though not claimed to be exact \cite{percoquest}.
One should also note that the jamming transition is always sharper than the
percolation transition, i.e. $\Delta_{jam} < \Delta_{perc}$. The length
$\Delta$ vanishes \cite{bookstauffer} as a power of the system size $L$.
One has
\begin{equation}
\Delta \sim L^{-1/\nu}
\end{equation} where $\sim$ means asymptotic proportionality. This allows
for the measure of the exponent $\nu$ for the correlation length $\xi$
which diverges at the critical point as $\xi \sim {|p_c - p|}^{-\nu}$. The
above relationship (2) means that the transition is sharper for larger
lattice sizes. As expected, the exponent value has been found to be
$\nu_{perc} = 1.35 \pm 0.02$ compatible with the $\nu_{perc} = 4/3$ value
known for $d=2$ percolation with isotropic particles. Another exponent
value is found for the jamming transition: $\nu_{jam} = 1.0 \pm 0.1$, a
value also reported in the work of Nakamura \cite{nakamura} about the
jamming of blocks. Both values $\nu_{perc}$ and $\nu_{jam}$ have been found
to be independent of the aspect ratio $a$ of the needles. It seems at first
that both critical phenomena are independent of each other since critical
exponents and thresholds are different. However, we will see herebelow that
a relationship between percolation and jamming thresholds nevertheless
exists!

The threshold values are slightly dependent on the system size $L$. The
following dependence \begin{equation} p_c(\infty) - p_c(L) \sim L^{-1/\nu}
\end{equation} has been established for percolation \cite{bookstauffer}.
The latter relationship allows us to extrapolate the threshold for an
infinite system $L \rightarrow \infty$. Table I summarizes some of the
extrapolated values of $p_c$ for both percolation and jamming phenomena for
typical values of $a$. When the aspect ratio $a$ of the needles increases,
both thresholds expectedly decrease. Figure 3 presents both $p_c^{perc}$
and $p_c^{jam}$ as a function of $a$. Curves are guides for the eye.

In order to search for a possible relationship between both thresholds if
any, we have calculated the ratio $p_c^{perc}/p_c^{jam}$ (see the third
column of Table I). Surprisingly, this latter ratio is found to be a
constant $0.62 \pm 0.01$ whatever the aspect ratio value $a$! Small
deviations from the 0.62 value are only observed for large $a$ values;
those deviations are certainly due to the particle/lattice finite size. The
universal ratio is unexpected and suggests that both critical phenomena are
intimately related, although we have found that the critical exponent
values are different. In fact, this result indicates that the percolation
cluster should be a fundamental cluster (like a skeleton) for the jamming
phase!

It should be noted that the scaling properties of the jamming phase are not
similar to those of dense systems otherwise the threshold $p_c^{jam}$ would
be independent of $a$. In the case of the jamming transition, a power law
behavior \begin{equation} p_c^{jam} \sim a^{-\delta} \end{equation} with an
exponent $\delta \approx 0.20$ has been proposed for off-lattice RSA of
rectangles \cite{ziff}. These authors \cite{ziff} have suggested that the
power law (4) indicates that the network of needles is fractal with a
dimension $2-\delta$ at $p_c^{jam}$ at scales below $a$. Moreover,
systematic deviations from Eq.(4) appear for large $a$ values. It seems
that both thresholds are decaying in a slower fashion than a power of $a$
(not shown in Figure 3). An empirical law will be proposed in the following.

Consider now the Nakamura work. The ratio between the thresholds
$p_c^{jam}$ for blocks and needles seems also to be a universal constant
(see the last column of Table I)! This remarkable result suggests that
there is also a relationship between needles and blocks for the jamming
phase. Of course, we have seen above that close packed needles form
clusters which can be seen as blocks of size $a \times a$.

Let us interpret our results by considering a coarse grained view of the
disordered systems. In a coarse grained view, i.e. at a scale larger than
$a$, the notion of anisotropy should disappear. Needles of size $a$ are
replaced by blocks of size $a \times a$. The threshold is then expected to
be that of classical square objects. Since the notion of anisotropy
disappears at larger scales, the fraction of occupied sites $p'$ in a
coarse grained view is larger than the true fraction of occupied sites $p$.
We propose the equation \begin{equation} p = p' \left[ 1 - \gamma {\left(
{a-1 \over a} \right)}^2 \right] \end{equation} where $\gamma$ is a
constant. The factor in the r.h.s. represents the fraction of free sites
remaining if two perpendicular needles are touching in any $a \times a$
``supersite". This term represents the loss of information when one looks
for the connectivity of the packing in a coarse grained view. This term
represents also the jamming phase since no more needles can be added in a
``supersite" which contains already 2 perpendicular touching needles. The
parameter $\gamma$ should depend only on geometrical aspects and should be
independent of $a$. Using the relationship (5), one expects that both
thresholds scale as \begin{equation} p_c(a) = C \left[ 1 - \gamma {\left(
{a-1 \over a} \right)}^2 \right] \end{equation} where $C$ is a constant.
The law (6) fits the data and is illustrated in Figure 4. The agreement is
quite remarkable. The coefficient $\gamma$ is found: $\gamma = 0.31 \pm
0.01$ in all cases, providing a constant ratio of thresholds. One should
note that the Eq.(6) is not valid for $a=1$ which is a particular
(isotropic needles!) point.

In the Nakamura case (blocks of size $a$), one has a loss of information
due to block edges. The empirical law of Eq.(6) is found to hold also in
that case! This good agreement suggests that our physical arguments are
appropriate to the study of packed anisotropic objects.

In summary, we have investigated two phenomena, i.e. percolation and
jamming, which have been independently studied up to now. We have found
that they are closely related in the case of anisotropic objects (needles).
We have interpreted this effect assuming a loss of information in a coarse
grained view. Fundamental laws have been obtained theoretically. Finite
size and shape effects seem to have been captured in a scaling law of
$(1-1/a)^2$ for needles.

\vskip 1.0cm {\noindent \large Acknowledgements} \vskip 0.6cm

NV thanks the FNRS for financial support. A critical reading of the
manuscript by R.Cloots and M.Ausloos is gratefully acknowledged.

\newpage

\begin{tabular}{|c|c|c|c|c|c|}
\hline
$a$ & \multicolumn{3}{c|} {needles} & {blocks} \\ & $p_c^{perc}$ &
$p_c^{jam}$ & $p_c^{perc} \over p_c^{jam}$ & $p_c^{jam}$ &
$p_c^{jam}(blocks) \over p_c^{jam}(needles)$ \\ \hline 2 & 0.562 & 0.906 &
0.621 & 0.749 & 0.827\\ 3 & 0.528 & 0.847 & 0.623 & 0.681 & 0.804\\ 4 &
0.504 & 0.811 & 0.622 & 0.646 & 0.797\\ 5 & 0.490 & 0.787 & 0.623 & 0.628 &
0.798\\ 6 & 0.479 & 0.770 & 0.622 & 0.620 & 0.805\\ 8 & 0.474 & 0.757 &
0.626 & 0.603 & 0.797\\ 10 & 0.467 & 0.741 & 0.630 & 0.593 & 0.800\\
\hline
\end{tabular}

\vskip 1.0cm Table I --- List of some thresholds $p_c$ for both percolation
and jamming phenomena for typical values of the needle aspect ratio $a$.
The ratio $p_c^{perc} / p_c^{jam}$ is also given in the fourth column. The
Nakamura thresholds for blocks in the jamming phase are given in the fifth
column. In that case, $a$ represents the block width. The last column gives
the ratios for blocks and needles thresholds $p_c^{jam}$.

\newpage {\noindent \large Figure captions}

\vskip 1.0cm {\noindent \bf Figure 1} --- (a) Percolation cluster of
non-overlapping needles of aspect ration $a=4$. (b) The largest cluster at
the jamming transition.

\vskip 1.0cm {\noindent \bf Figure 2} --- Jamming phase for $a=20$ needles.
The lattice size is $500 \times 500$. Vertical and horizontal orientations
are represented by different grey levels.

\vskip 1.0cm {\noindent \bf Figure 3} --- The thresholds $p_c^{perc}$ and
$p_c^{jam}$ as a function of the needle aspect ratio $a$. The nakamura
results are also illustrated. The curves are only guides for the eye.

\vskip 1.0cm {\noindent \bf Figure 4} --- The thresholds $p_c^{perc}$ and
$p_c^{jam}$ as a function of $(a-1)^2/a^2$. The Nakamura results are also
illustrated. The straight lines are fits using Eq.(6).

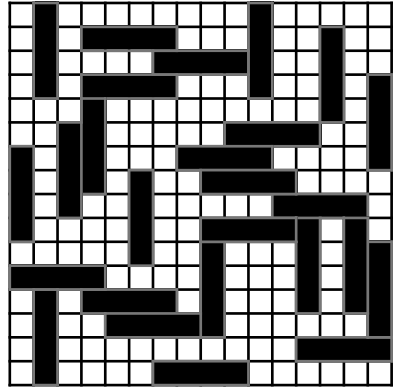
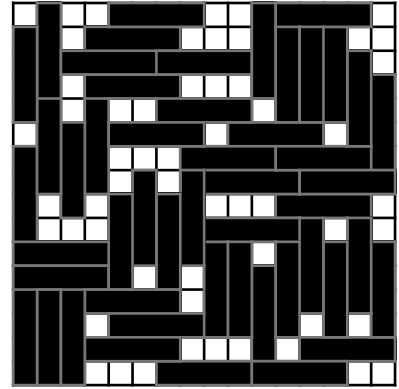
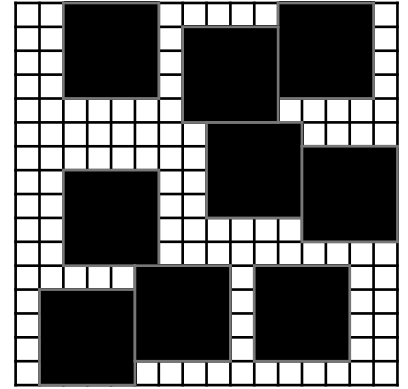
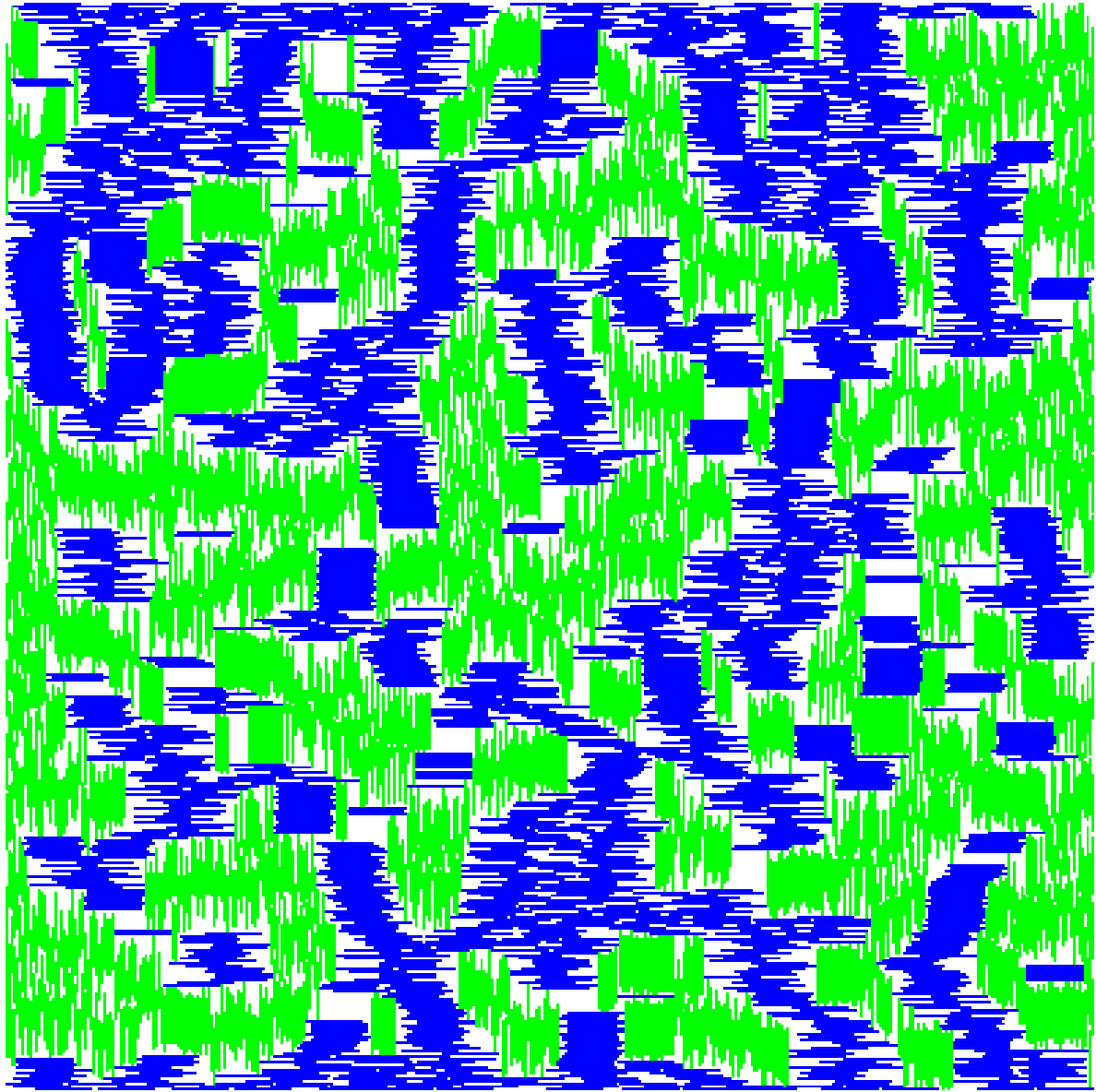
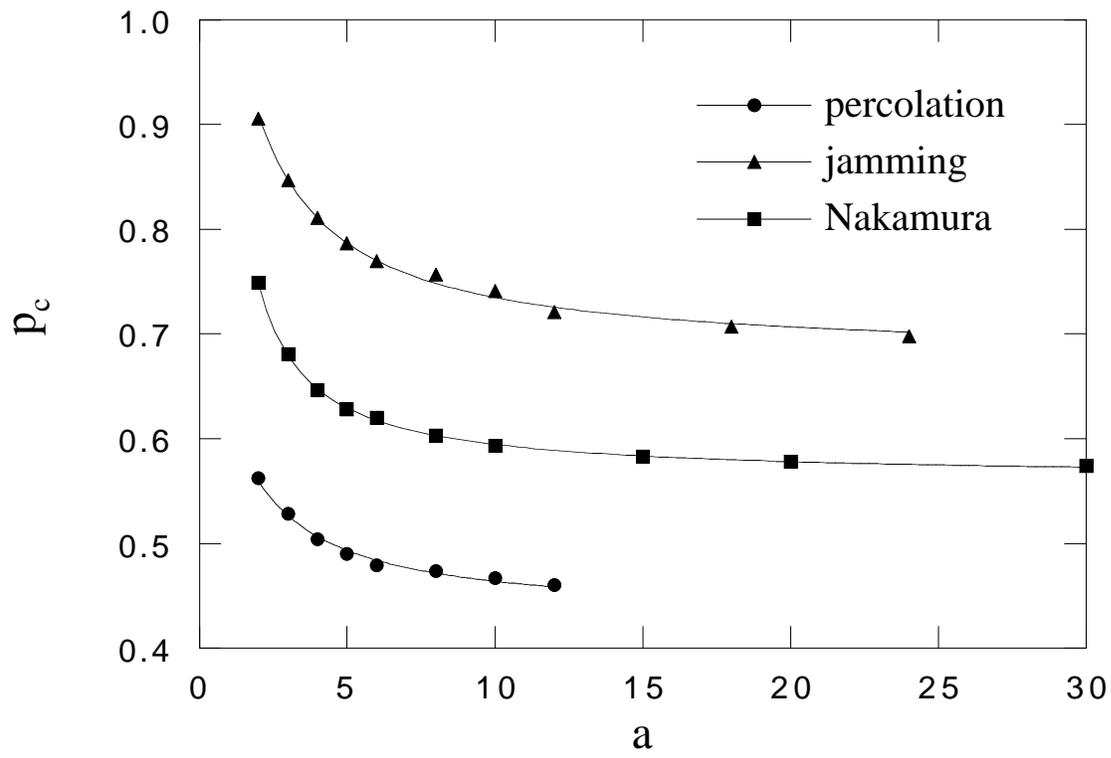
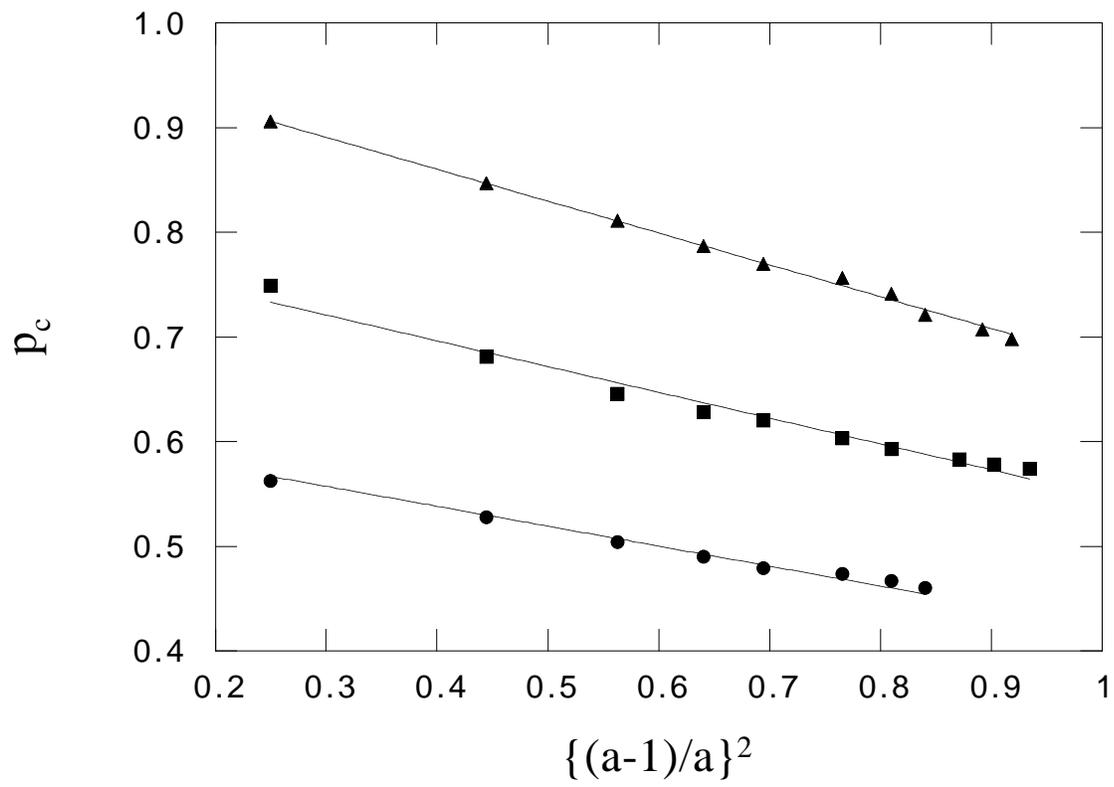
\end{document}